\def\e{\varepsilon}
\newcommand{\lsim}{\raisebox{-.020in}{$\stackrel{<}{{\scriptstyle
\sim}}$}}

\def\sqr#1#2{{\vcenter{\vbox{\hrule height.#2pt
\hbox{\vrule width.#2pt height#1pt \kern#1pt
\vrule width.#2pt}
\hrule height.#2pt}}}}

\documentstyle[aps,multicol]{revtex}

\begin{document}

\begin{multicols}{2}
\columnseprule 0pt
\narrowtext

\noindent{\Large Comment on ``Chain Length Scaling of Protein Folding Time''}

\bigskip

In a recent Letter, Gutin, Abkevich, and Shakhnovich [1] reported on a 
series of dynamical Monte Carlo simulations on lattice
models of proteins. Based on these highly simplified models, 
they found that four different
potential energies lead to four different folding time scales
$\tau_f$, where $\tau_f$ scales with chain length as $N^\lambda$
(see, also, Refs. [2-4]), with 
$\lambda$ ranging from 2.7 to 6.0. However, because of the lack of
microscopic models of protein folding dynamics,
the interpretation and 
origin of the data have
remained somewhat speculative.
It is the
purpose of this Comment to point out that the application
of a simple ``mesoscopic'' model of protein folding [2] provides
a quantitative interpretation of the data, as well as
a major qualitative difference with Ref. [1]. 

The main features of the theory [2] are as follows. Consistent with
the heuristic arguments regarding the entropic cost
of loop formation [1,3], we model the acquisition of native-like features
by considering the full set of 
loops or links of an ideal chain of $N$ monomers, and
a target state defined by a unique set of $N/2$ ``native-like'' links.
Folding is tuned by a {\it microscopic} frustration parameter
$\Delta$, defined as the contact energy ratio $\delta/\e$ between
random and native-like links, respectively.
For a finite chain,
the model has an effective
folding transition temperature $T_f(N)\sim (1-\Delta)/\ln N$, between random
conformations and native-like structures.
At $T_f(N)$, a
bimodal distribution on the number of non-native contacts is observed,
whereas away from $T_f(N)$ distributions
are unimodal. We point out that the latter is commonly associated
with ``cooperativity'' [1]. However, we predict 
the size of the critical folding nucleus to be
proportional to the size of the system. The same conclusion 
is found based on exact enumerations of self-avoiding walks [5].
In agreement with the ideal ``minimally frustrated'' Go model 
$\lambda^{Go} \simeq 2.7$ [1],
optimal folding to fully equilibrated native-like structures takes place in
an entropic dominated
time scale $\tau_f \sim N^{\lambda_0}$, where $\lambda_0\simeq 3$. The striking
prediction of the theory is that at the 
transition $T_f(N)$ ---which, in all likelihood, corresponds
to the ``optimal'' temperature estimated using mean 
first passage time simulations
[1]--- $\tau_f(T_f)\sim N^\lambda$, where
$\lambda\simeq 4$ (see Fig. 1), the same behavior {\it now}
being confirmed by simulations [1] on specially designed fast folding
sequences!

This scaling breaks down when the frustration limited folding time scale
$\tau_\Delta\sim  N^{\lambda_0}\exp(a\e\Delta/T)$ becomes larger than $\tau_f$,
$a\sim O(1)$ being
a constant.
These scaling forms predict that, for small
enough $\Delta \,\lsim\, 1/3$, proteins fold fast in a
unique range of temperatures roughly
independent of size and structural specificities, i.e. $T_\Delta(N)<T<T_f(N)$,
where $T_\Delta$ is a dynamically determined transition
temperature at which $\tau_f\simeq\tau_\Delta$. Below
$T_\Delta$ the relaxation is slow, following an Arrhenius law [1] 
(higher energy barriers not included here may also be important in
this regime).
If $T_f(N)<T_\Delta(N)$ frustration dominates altogether,
and folding can be extremely
slow [2], with
\begin{equation}
\tau_f(T_f)\sim N^{\lambda_0+2\Delta/(1-\Delta)}.
\end{equation}

Simulations have shown that different potential energies have
different landscapes with different degrees of frustration [1,4].
The model averages 
frustration into a single {\it model dependent} parameter $\Delta$.
As sketched in Fig. 3 of Ref. [2], random or frustration dominated
sequences with $\Delta\,\lsim\,1/2$ will not be able to fold in a 
physically relevant time scale.
Equating $\lambda^{RAN}\simeq 6$ with the exponent in (1), we find $\Delta^{RAN}\simeq
3/5$, which is quite appropriate for  a random selection of contact
energies (cf. [2]).

It has been claimed that these results can be understood in the
context of a first-order-like folding transition [1]. However,
it is well known that near a first-order transition different
relaxation times grow in different ways.
In particular, these authors have also argued about the existence of a 
macroscopic barrier between native and non-native states 
(``cooperativity''). Thus, if this barrier needs to be
crossed over,
the relaxation must then be exponential in $N$! On the contrary,
we find an effective second order folding transition, where
at $T_f(N)$ any typical relaxation time must grow as a  power of $N$.
In summary, 
the data in [1] is accounted for by [2], giving
further credence to the cross-linking
model, and emphasizing 
the role of loop formation in folding kinetics. 

Support from Fondecyt Nr. 3940016 is acknowledged.
\bigskip

\noindent{Carlos J. Camacho}
\par Facultad de F\'\i sica, Universidad Cat\'olica de Chile
\par Casilla 306, Santiago 22, Chile

\bigskip

\noindent{PACS numbers: 87.15.By, 82.20.Db, 05.70.Fh, 64.60.Cn}

\bigskip

\noindent [1] A.M. Gutin {\it et al.}, Phys. Rev. Lett. {\bf 77},
5433 (1996).\hfill\break
[2] C.J. Camacho, Phys. Rev. Lett. {\bf 77}, 2324 (1996).\hfill\break
[3] D. Thirumalai, 
J. Phys. (Paris) {\bf 5}, 1457 (1995).\hfill\break
[4] C.J. Camacho and D. Thirumalai,  Proc. Natl. Acad. Sci. USA
{\bf 90}, 6369 (1993).\hfill\break
[5] C.J. Camacho and T. Schanke, cond-mat/9604174. \hfill\break

\begin{figure}
\caption{Folding time scale at the transition as a function of size $N$.
The slopes
correspond to $3.8\pm.25$ ($\Delta=0$) and $4.3\pm .2$ ($\Delta=1/3$) [2].
For details of master equation see [2].
}
\end{figure}

\vfill\break
\end{multicols}

\end{document}